\documentclass[12pt]{article}
\usepackage{enumerate}
\usepackage{graphics}
\usepackage{amsmath}
\usepackage{amssymb}
\usepackage{array}
\usepackage{geometry}
\usepackage{graphicx}
\usepackage{enumerate}
\usepackage{lscape,graphicx}
\usepackage{fancyhdr}
\usepackage{wasysym}
\usepackage{chapterbib}

\begin{document}

{\bf{ \Large\noindent  An Overview of the Pathway Idea in Statistical and Physical Sciences }}
\vskip.3cm
\noindent
\begin{center}
\textbf{Nicy Sebastian} \\
Indian Statistical Institute, Chennai Centre, Taramani, Chennai - 600113\\
nicy@isichennai.res.in\\

\textbf{Dhannya P. Joseph}\\
Kuriakose Elias College, Mannanam P.O, Kottayam, Kerala - 686 561\\
 dhannyapj@gmail.com\\
\textbf{Seema S. Nair }\\
Centre for Mathematical Sciences,
Arunapuram P.O.,
 Palai, Kerala - 686 574\\
 seema.cms@gmail.com \\
\end{center}
\vskip .4cm
\noindent
{\bf Summary}
\vskip .2cm
{\small{\bf{
 Pathway idea is a switching mechanism by which one can go from one functional form to another, and to yet another. It is shown that through a parameter $\alpha$, called the pathway parameter, one can connect generalized type-1 beta family of densities, generalized type-2 beta family of densities, and generalized gamma family of densities, in the scalar as well as the matrix cases, also in the real and complex domains.
It is shown that when the model is applied to physical situations then the current hot topics of Tsallis statistics and superstatistics in statistical mechanics become special cases of the pathway model, and the model is capable of capturing many stable situations as well as the unstable or chaotic neighborhoods of the stable situations and transitional stages. The pathway model is shown to be connected to generalized information measures or entropies, power law, likelihood ratio criterion or $\lambda-$criterion in multivariate statistical analysis, generalized Dirichlet densities, fractional calculus, Mittag-Leffler stochastic process, Kr\"{a}tzel integral in applied analysis, and many other topics in different disciplines. The pathway model enables one to extend the current results on quadratic and bilinear forms, when the samples come from Gaussian populations, to wider classes of populations.
}}
\vskip .3cm
\noindent
{\it Keywords}: Pathway model; beta family; generalized gamma; $\lambda$-criterion; Dirichlet densities; $H$- function; quadratic forms.}\\
\vskip .6cm

\noindent
{\bf 1 \hskip.3cm Introduction}
\vskip.2cm
\par
The pathway idea was originally prepared by Mathai in the 1970's in connection with population models, and later rephrased and extended Mathai (2005) to cover scalar as well as matrix cases as made suitable for modelling data from statistical and physical situations.
For practical purposes of analyzing data of physical experiments and in building
up models in statistics, we frequently select a member from a
parametric family of distributions.
But it is often found that the model requires a distribution with
a thicker or thinner tail than the ones available from the parametric family, or
a situation of right tail cut-off. The experimental data reveal that the underlying distribution is in between
two parametric families of distributions.
 In order to create a pathway from one functional form to another, a
pathway parameter is introduced and a pathway model is created in Mathai
(2005). The main idea behind the derivation of this model is the switching
 properties of going from one family of functions to another and yet another family of functions.
 The model enables one to proceed from a generalized type-1 beta
model to a generalized type-2 beta model to a generalized gamma model
when the variable is restricted to be positive. Thus the pathway parameter $\alpha$ takes one to three different functional forms. This is the distributional pathway. More families are available
when the variable is allowed to vary over the real line. Mathai (2005) deals
mainly with rectangular matrix-variate distributions and the scalar case is
a particular case there. For the real scalar case the pathway model is the
following:
\begin{equation}\label{pathwayless1}
f_1(x)=c_1x^{\gamma-1}[1-a(1-\alpha)x^{\delta}]^{\frac{\eta}{1-\alpha}},
\end{equation}
$a>0,~\delta>0,~1-a(1-\alpha)x^{\delta}>0,~\gamma>0,~\eta>0$
where $c_1=\frac{\delta
(a(1-\alpha))^{\frac{\gamma}{\delta}}\Gamma(\frac{\eta}{1-\alpha}+1+\frac{\gamma}{\delta})}
{\Gamma(\frac{\gamma}{\delta})\Gamma(\frac{\eta}{1-\alpha}+ 1)},$ is the normalizing constant if a statistical density is needed and
$\alpha$ is the pathway parameter. The model in (\ref{pathwayless1}) exactly $c_1$ can be used for modeling physical situations. For $\alpha < 1$ the model remains as a generalized
type-1 beta model in the real case. 
Other cases available are the regular
type-1 beta density, Pareto density, power function, triangular and related
models. Observe that (1) is a model with the right tail cut off. When $\alpha> 1$
we may write $1 - \alpha = -(\alpha - 1),~ \alpha > 1$ so that $f(x)$ assumes the form,
\begin{equation}\label{pathwaygreater1}
f_2(x) = c_2x^{\gamma-1}
[1 + a(\alpha - 1)x^\delta]^{-\frac{\eta}{\alpha-1}}
 ,~ x > 0, \end{equation}
which is a generalized type-2 beta model for real $x$ and $c_2=\frac{\delta
(a(\alpha-1))^{\frac{\gamma}{\delta}}\Gamma(\frac{\eta}{\alpha-1})}
{\Gamma(\frac{\gamma}{\delta})\Gamma(\frac{\eta}{\alpha-1}-
\frac{\gamma}{\delta})},$ is the normalizing constant, if a statistical density is required.  Beck and Cohen's superstatistics
belong to this case (\ref{pathwaygreater1}) (Beck $\&$ Cohen, 2003; Beck, 2006). Again, dozens of published papers are available on the topic superstatistics in astrophysics. For
$\gamma= 1,~ a = 1, \delta = 1$ we have Tsallis statistics for $\alpha > 1 $ from (\ref{pathwaygreater1}). Other
standard distributions coming from this model are the regular type-2 beta,
the F-distribution, L$\acute{e}$vi models and related models. When $\alpha\rightarrow 1,$ the forms
in (\ref{pathwayless1}) and (\ref{pathwaygreater1}) reduce to
\begin{equation}\label{gamma}
f(x) = cx^{\gamma-1}{\rm e^{-b x^{\delta}}}
,~ x >0,~b=a\eta,
\end{equation}
where $c=\frac{\delta
b^{\frac{\gamma}{\delta}}}{\Gamma(\frac{\gamma}{\delta})},$ is the normalizing constant.
This includes generalized gamma, gamma, exponential, chisquare, Weibull,
Maxwell-Boltzmann, Rayleigh, and related models (Mathai, 1993a; Honerkamp,
1994). If $x$ is replaced by $|x|$ in (\ref{pathwayless1}) then more families of distributions
are covered in (\ref{pathwayless1}). The behavior of the pathway model for various values of the pathway parameter $\alpha$ can be seen from the following figures. \\

\begin{figure}
\begin{center}
~~~~~ \resizebox{7cm}{5cm}{\includegraphics{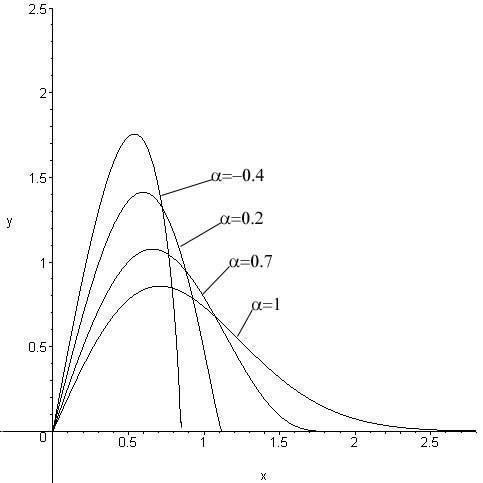}}
~~~~ \resizebox{7cm}{5cm}{\includegraphics{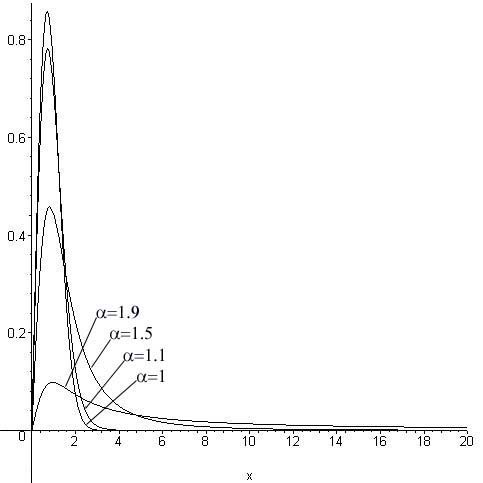}}\\
\caption {\small (a)The graph of $f_{1}(x)$, for $\gamma = \eta = a = 1,~
\delta=2$ and for various values of $\alpha$.
\hskip1.5cm (b)The graph of $f_{2}(x)$, for $\gamma= \eta= a =1,
\delta=2$ and for various values of $\alpha$.\label{ppplot1}}
\end{center}
\end{figure}

%
From the Figure \ref{ppplot1}(a) we can see that, as $\alpha$ moves away from $1$
the function $f_{2}(x)$ moves away from the origin and it
becomes thicker tailed and less peaked. From the path created by
$\alpha$ we note that we obtain densities with thicker or thinner
tail compared to generalized gamma density. From Figure \ref{ppplot1}(b) we can see
that when $\alpha$ moves from $-\infty$ to $1$, the curve becomes thicker
tailed and less peaked, see Joseph (2009), Haubold et al. (2010). Note that $\alpha$ is the most important parameter here for enabling one to more than one family of functions to another family. The other parameters are the usual parameters within each family of functions. The following is a list of some particular cases and the transformations are listed to go from the extended versions to the regular cases.
{\scriptsize{\begin{center}
\begin{tabular}{|c|c|}\hline
           ~&~\\
        $\alpha=1,~\gamma=1,~a=1,~\delta=1$& Gaussian or normal density for $\infty<x<\infty$ \\
        \hline
          ~&~\\
        $\alpha=1,\gamma-1=\frac{3}{4},~a=1,~\delta=1$ & Maxwell-Boltzmann density \\
                \hline
          ~&~\\
$\alpha=1,~\gamma-1=\frac{1}{2},~a=1,~\delta=1$& Rayleigh density\\
 \hline
          ~&~\\
         $\alpha=1,~\gamma=\frac{n}{2},~a=1,~\delta=1$& Hermert density\\
        \hline
        ~&~\\
        $\alpha=0,~\gamma=1,~\eta=1,~\delta=1$& U-shaped density\\
         \hline
                   ~&~\\
         $\alpha=2,~\gamma=1,\eta=\frac{\nu+1}{2}~a=\frac{1}{\nu},~\delta=1$& Student-t for $\nu$ degrees of freedom,~$-\infty<x<\infty$\\
         \hline

           ~&~\\
         $\alpha=2,~\eta=1,~a=1,~\delta=1$& Caushy density for $-\infty<x<\infty$\\
               \hline
~&~\\
         $\alpha<1,~a(1-\alpha)=1,~\delta=1$& Standard type-1 beta density\\
                 \hline
                        \end{tabular}
\end{center}}}
{\scriptsize{\begin{center}
\begin{tabular}{|c|c|}\hline
  ~&~\\
         $\alpha>1,~a(1-\alpha)=1,~\delta=1$& Standard type-2 beta density\\
         \hline
                           ~&~\\
         $\gamma-1=\frac{1}{2},~\eta=1,~a=1,~\delta=1$& Tsallis statistics in Astrophysics,  \\ ~&Power law, $q$-binomial density\\
           \hline
            ~&~\\
         $\alpha=0,~\gamma-1=\frac{1}{2},~\eta=1,~\delta=1$& Triangular density\\
         \hline

          ~&~\\
         $\alpha=2,~\gamma-\frac{1}{2}=\frac{m}{2},a=\frac{m}{n}~\eta=\frac{m+1}{2},~\delta=1$& F-density\\
         \hline
          ~&~\\
         $\alpha=1,~\gamma-1=\frac{1}{2},~a=1,~\eta=\frac{mg}{KT},~\delta=1$& Helley's density in physics\\
              \hline
                        ~&~\\
         $\alpha=1,~a=1,~\delta=1$& Gamma density\\
         \hline
          ~&~\\
         $\alpha=1,~a=1,~\gamma-\frac{1}{2}=\frac{\nu}{2},~\eta=\frac{1}{2},~\delta=1$& Chisquare density for $\nu$ degrees of freedom\\
         \hline
          ~&~\\
         $\alpha=1,~a=1~\gamma-1=\frac{1}{2},~\delta=1$& Exponential density (Laplace density \\
         ~&with $x=|z|,~-\infty<z<\infty$)\\
         \hline
          ~&~\\
         $\alpha=1,~a=1$& Generalized gamma density\\
         \hline
                   ~&~\\
         $\alpha=1,~a=1,~\gamma-1=\frac{1}{2}$& Weibull density\\
         \hline
          ~&~\\
         $\alpha=2,~a=1,~\gamma-1=\frac{1}{2},~\eta=2,~\delta=1,~x={\rm e}^y$& Logistic density for $-\infty<y<\infty$\\
          \hline
~&~\\
         $\alpha=2,~a=1,~\gamma=1,~\eta=1,~\delta=1,~x={\rm e}^{\epsilon+\mu y},\epsilon \neq 0,~\mu>0,$& Fermi-Dirac density,~$0\leq y<\infty$\\
         \hline
\end{tabular}
\end{center}}}

 \vskip.3cm
\noindent
\emph{1.1 \hskip.3cm Pathway model from Mathai's entropy measure}
\vskip.3cm
In physical situations when an appropriate density is selected, one
procedure is the maximization of entropy\index{entropy}. Mathai $\&$ Rathie (1975)
consider various generalizations of Shannon\index{Shannon entropy}
entropy\index{entropy} measure and describe various properties
including additivity, characterization theorem etc. Mathai $\&$ Haubold (2007a)
introduced a new generalized entropy\index{entropy} measure which is
a generalization of the Shannon \index{Shannon entropy}
entropy\index{entropy} measure. For a multinomial population $P=
(p_1, \ldots, p_k), ~p_i\geq 0,~ i=1, \ldots, k,~p_1+p_2+\cdots+p_k
=1$,  the Mathai's\index{Mathai's Entropy} entropy\index{entropy}
measure is given by the relation {\small
$$M_{k,\alpha }(P)= \frac{\displaystyle\sum_{i=1}^k p_i^{2-\alpha }-1}{\alpha
-1}, \; \alpha \neq 1,\;-\infty <\alpha <2.~~~~~~\text{(discrete case)}$$}
{\small
$$M_{\alpha }(f)= \frac{1}{\alpha-1}\left[\int_{-\infty}^\infty[f(x)]^{2-\alpha}{\rm d}x-1\right],~\alpha \neq1,~\alpha<2~~~~~~\text{(continuous case)}$$}

\noindent
By optimizing Mathai's entropy\index{entropy} measure, one can arrive at pathway model of Mathai (2005), which consists of  many of the standard
distributions in statistical literature as special cases. For fixed $\alpha$, consider the optimization of $M_\alpha(f)$, which implies optimization of $\int_x[f(x)]^{2-\alpha}{\rm d}x$, subject to the following conditions:
\begin{enumerate}[{(i)}]
\item $f(x)\geq 0,~\text{for all}~x$
\item $\int_x f(x){\rm d}x < \infty$
\item $\int_x x^{\rho(1-\alpha)}f(x){\rm d}x=\text{fixed for all}~f$
\item $\int_x x^{\rho(1-\alpha)+\delta}f(x){\rm d}x=\text{fixed for all}~f, \text{where}~\rho~ \text{and}~\delta ~\text{are fixed parameters}$
\end{enumerate}
By using calculus of variation, one can obtain the Euler equation as
\begin{eqnarray}
&&\frac{\partial}{\partial f}[f^{2-\alpha}-\lambda_1x^{\rho(1-\alpha)}f+\lambda_2x^{\rho(1-\alpha)+\delta}f]=0\nonumber\\
&&\Rightarrow (2-\alpha)f^{1-\alpha}=\lambda_1x^{\rho(1-\alpha)}[1-\frac{\lambda_2}{\lambda_1}x^\delta],~\alpha\neq 1,2\nonumber\\
&&\Rightarrow f_1=c_1x^\rho[1-a(1-\alpha)x^\delta]^{\frac{1}{1-\alpha}}
\end{eqnarray}
for $\frac{\lambda_2}{\lambda_1}=a(1-\alpha)$ for some $a>0$. For more details the reader may refer to the papers of Mathai $\&$ Haubold (2008, 2007b).\\
\par
When $\alpha \rightarrow 1$,  the Mathai's entropy measure $M_{\alpha}(f)$ goes to
the Shannon \index{Shannon entropy} entropy\index{entropy} measure
and this is a variant of
 Havrda-Charv\'{a}t entropy\index{entropy}, and the variant form therein is Tsallis
 entropy\index{entropy}. Then when $\alpha$ increases from 1, $M_{\alpha }(f)$ moves away from Shannon entropy. Thus $\alpha$ creates a pathway moving from one function to another, through the generalized entropy also. This is the entropic pathway. One can derive  Tsallis statistics and superstatistics (Beck (2006) $\&$ Beck and Cohen (2003)) by using  Mathai's\index{Mathai's Entropy} entropy.\\

 The pathway parameter $\alpha$ offers the differential pathway also. Let us consider
 \begin{eqnarray}\label{differential}
 g(x)&=&\frac{f_1(x)}{c_1}=x^\beta[1-a(1-\alpha)x^\delta]^{\frac{1}{1-\alpha}},~x>0, \beta=\gamma-1, \eta=1\nonumber\\
 \frac{{\rm d}}{{\rm d}x}g(x)&=&\frac{\beta}{x}g(x)-a\delta x^{\delta-1+(1-\alpha)\gamma}[g(x)]^\alpha\nonumber\\
 &=&-a[g(x)]^\alpha~~\text{for}~~\beta=0,~\delta=1
  \end{eqnarray}
  This is the power law. When $\eta=1$  then the differential equation satisfied by $g_3=\frac{f}{c}$ of (\ref{gamma}) is given by
\begin{eqnarray}\label{differential1}
 \frac{{\rm d}}{{\rm d}x}g_3(x)&=&\frac{\beta}{x}-a\delta x^{\delta-1}g_3(x)\nonumber\\
 &=&-a[g_3(x)]~~\text{for}~~\beta=\gamma-1=0,~\delta=1
  \end{eqnarray}
Thus when $\alpha$ moves to 1 the differential pathway is from the power law in (\ref{differential}) to the maxwell-Boltzmann in (\ref{differential1}).
\vskip.3cm
\noindent
\emph{1.2 \hskip.3cm Laplacian density and stochastic
processes}
\vskip.3cm
 The real scalar case of the pathway model in
(\ref{pathwayless1}), when  $x$ is replaced by $|x|$ and $\alpha\rightarrow 1$, takes the form
\begin{equation}\label{|x|}
f_3(x)=c_3|x|^{\gamma-1}{\rm e}^{-a|x|^{\delta}}, -\infty <x<\infty, a>0.
\end{equation}
The density in (\ref{|x|}) for $\gamma=1, \delta=1$ is the simple Laplace
density. For $\gamma=1$ we have the symmetric Laplace density. A
general Laplace density is associated with the concept of
Laplacianness of quadratic and bilinear forms. For the concept of
Laplacianness of bilinear forms, corresponding to the chisquaredness
of quadratic forms, and for other details see Mathai (1993a) and
Mathai et al. (1995). Laplace density is also
connected to input-output type models. Such models can describe many
of the phenomena in nature. When two particles react with each other
and energy is produced, part of it may be consumed or converted or
lost and what is usually measured is the residual effect. The water
storage in a dam is the residual effect of the water flowing into
the dam minus the amount taken out of the dam. Grain storage in a
sylo is the input minus the grain taken out. It is shown in Mathai
(1993b) that when we have gamma type input and gamma type output the
residual part $z=x-y,~ x=$ input variable, $y= $ output variable, then
the special cases of the density of $z$ is a Laplace density. In
this case one can also obtain the asymmetric Laplace and generalized
Laplace densities, which are currently used very frequently in
stochastic processes, as special cases of the input-output model.
Some aspects of the matrix version of the input-output model is also
described in Mathai (1993b).
\vskip.3cm
\noindent
\emph{1.3 \hskip.3cm Mittag-Leffler density and processes}
\vskip.3cm
Recently there is renewed interest in Mittag-Leffler
 function as a model in many applied areas due to many reasons, one
 being that it gives a thicker tail compared to the exponential
 model. Fractional differential equations often lead to Mittag-Leffler
 functions and their generalizations as solutions, especially when dealing with fractional
 equations in reaction-diffusion problems. A large number of such
 situations are illustrated in Mathai \& Haubold (2008), and Mathai et al. (2010), Mathai (2010).  The Mittag-Leffler density, associated with a
3-parameter Mittag-Leffler function is the following:
\begin{equation}
f(x)=\frac{x^{\alpha\beta-1}}{\delta^{\beta}}\sum_{k=0}^{\infty}\frac{(\beta)_k}{k!}
 \frac{(-x^{\alpha})^k}{\delta^k\Gamma(\alpha
 k+\alpha\beta)},0\le x<\infty,\delta>0,\beta>0.
 \end{equation}
 It has the Laplace transform
 \begin{equation}
L_f(t)=[1+\delta t^{\alpha}]^{-\beta},~ 1+\delta t^{\alpha}>0
\end{equation}
 If $\delta$ is replaced by $\delta(q-1)$ and $\beta$ by
 $\frac{\beta}{(q-1)}, q>1$ and if we consider $q$ approaching to $1$ then
 we have
\begin{equation}
 \lim_{q\rightarrow 1} L_f(t)=\lim_{q\rightarrow
 1}[1+\delta(q-1)t^{\alpha}]^{-{{\beta}\over{q-1}}}={\rm
 e}^{-\delta\beta t^{\alpha}}.
 \end{equation}
 But this is the Laplace transform of a constant multiple of a positive L\'evy variable
 with parameter $\alpha,0<\alpha\le 1$, with the multiplicative
 constant being $(\delta\beta)^{{1}\over{\alpha}}$, and thus the limiting form of
 a Mittag-Leffler distribution is a L\'evy distribution. A connection of pathway model
  to Mittag-Leffler function is given in Mathai \& Haubold (2010).
   There is vast literature on Mittag-Leffler stochastic processes, see for example Shanoja (2010, (Thesis)) and Pillai (1990).
\vskip.3cm
\noindent
 \emph{1.4 \hskip.3cm Laplace transform of the pathway model}
 \vskip.3cm
Let $L_{f_{2}}(t)$ be the Laplace transform of the pathway model
$f_{2}(x)$ of (\ref{pathwaygreater1}). That is
\begin{eqnarray}\label{laplaceext2}
L_{f_{2}}(t)&=&c_{2}\int_{0}^{\infty}{\rm
e}^{-tx}x^{\gamma-1}[1+a(\alpha-1)x^\delta]^{-{\frac{\eta}
{\alpha-1}}}{\rm
d}x,~a>0,~b>0,~\delta>0,\nonumber\\&&~~~~~~~~~~~~~~~~~~~~~~~~~~~~
~~~~~~~~~~~~~~~~~~~~~~~~~~~~~~~~~~~~~~~~~\eta>0,\alpha>1.
\end{eqnarray}
Here the integrand can be taken as a product of positive integrable functions and
then we can apply Mellin transform and inverse Mellin transform technique to evaluate the above integral.
The integral in  (\ref{laplaceext2}) can be looked upon as the Mellin convolution of
exponential density and superstatistics. Let us transform
$x_1$ and $x_2$ to $u=\frac{x_1}{x_2}$ and $v=x_2$, then we can see that the marginal density of
$u$ obtained is actually the Laplace transform that we want to evaluate.
Since the density is unique, in whatever way we evaluate the density we should get
the same function or the functions will be identical. We can evaluate the density
of $u$ by the method of inverse Mellin transform, see Joseph (2009). Comparing the density obtained in these
two different methods, we will get the Laplace transform of the pathway model given in equation (\ref{pathwaygreater1}) as
an $H$-function

\begin{eqnarray}\label{laplaceext1final}
L_{f_{2}}(t)=\frac{1}{\Gamma(\frac{\gamma}{\delta})\Gamma(\frac{\eta}{\alpha-1}-
\frac{\gamma}{\delta})}H_{1,2}^{2,1}\biggl[\frac{t}{a^{\frac{1}{\delta}}
(\alpha-1)^ {\frac{1}{\delta}}}
 \bigg|_{(0,1),(\frac{\eta}{\alpha-1}-\frac{\gamma}{\delta}
-\frac{1}{\delta},{\frac{1}{\delta}})}^{(1-\frac{\gamma}{\delta},
\frac{1}{\delta})}\biggr],
\end{eqnarray}
for
$\Re(\gamma)>0,~\Re(\frac{\eta}{\alpha-1}-\frac{\gamma}{\delta})>0,~\alpha>1,$
where $H-$ function is defined as
\begin{eqnarray}\label{hfunction}
H_{p,q}^{m,n}\bigl[z\big|_{(b_1,\beta_1),,...,(b_q,\beta_q)}^
{(a_1,\alpha_1),...,(a_p,\alpha_p)}\bigr]&=&\frac{1}{2\pi
i}\int_L\phi (s)~z^{-s}{\rm d}s,
\end{eqnarray}
where
\begin{eqnarray}
 \phi (s)&=&\frac{\bigl\{\prod _{j=1}^m \Gamma
(b_j+ \beta_js)\bigr\}~\bigl\{\prod _{j=1}^n\Gamma (1-a_j-
\alpha_js)\bigr\}}
 {\bigl\{\prod _{j=m+1}^q\Gamma (1-b_j-\beta_j s)\bigr\}~
\bigl\{ \prod _{j=n+1}^p \Gamma (a_j+\alpha_js)\bigr\}}\nonumber,
\end{eqnarray}
where $\alpha_j,~j=1,2,...,p$ and $\beta_j,~j=1,2,...,q$
 are real positive numbers, $a_j,~j=1,2,...,p$ and $b_j,~j=1,2,...,q$ are complex
numbers, $L$ is a contour separating the poles of $\Gamma
(b_j+\beta_js),~j=1,2,...,m$ from those of $\Gamma
(1-a_j-\alpha_js),~j=1,2,...,n$. When $\alpha_1=1=\cdots=\alpha_p=\beta_1=1=\cdots=\beta_q$, then the $H-$function reduces to Meijer's $G-$ function, for more details see Mathai et al. (2010). In a similar way we can evaluate the Laplace transform of the
pathway model for $\alpha<1$ and is given by
\begin{eqnarray}\label{laplaceext2final}
L_{f_{1}}(t)&=&c_{1}\int_{0}^{\bigl[\frac{1}{a(1-\alpha)}\bigr]^{\frac{1}
{\delta}}}{\rm
e}^{-tx}x^{\gamma_1}[1-a(1-\alpha)x^\delta]^{{\frac{\eta}
{1-\alpha}}}{\rm
d}x\nonumber\\&=&\frac{\Gamma(1+\frac{\eta}{1-\alpha}+\frac{\gamma}{\delta})}
{\Gamma(\frac{\gamma}{\delta})}H_{1,2}^{1,1}\biggl[\frac{t}{a^{\frac{1}{\delta}}
(1-\alpha)^ {\frac{1}{\delta}}}
 \bigg|_{(0,1),(-\frac{\eta}{1-\alpha}-\frac{\gamma}{\delta}
,{\frac{1}{\delta}})}^{(1-\frac{\gamma}{\delta},
\frac{1}{\delta})}\biggr],~~\Re(\gamma)>0
\end{eqnarray}
\noindent{\sc{Theorem 1. }}{\it{For
$\Re(\gamma)>0,~\Re(\frac{\eta}{\alpha-1}-\frac{\gamma}{\delta})>0,~\delta>0,~\eta>0,~x>0,~\alpha>1,$
the Laplace transform (or the corresponding moment generating
function) of the pathway model of the form $f_{2}(x),$
given in (\ref{laplaceext1final}) goes to the Laplace transform
(moment generating function) of the generalized gamma density given
in (\ref{laplacegenfinal}) when $\alpha\rightarrow1_{+}$.}}\\\\
\noindent{\sc{Theorem 2. }}{\it{For
$\Re(\gamma)>0,~\delta>0,~\eta>0,~x>0,~\alpha<1,$ the Laplace
transform (or the corresponding moment generating function) of the
pathway model of the form $f_{2}(x),$ given in
(\ref{laplaceext2final}) goes to the Laplace transform (moment
generating function) of the generalized gamma density given in
(\ref{laplacegenfinal}) when $\alpha\rightarrow1_{-}$.}\\\\}
The limiting case is given by
\begin{eqnarray}\label{laplacegenfinal}
L_{f}(t)&=&c\int_{0}^{\infty}e^{-tx}x^{\gamma}e^{-b x^{\delta}}{\rm
d}x\nonumber\\&=&\frac{1}{\Gamma(\frac{\gamma}{\delta})}
H_{1,1}^{1,1}\biggl[\frac{t}{b^{\frac{1}{\delta}}}
 \bigg|_{(0,1)}^{(1-\frac{\gamma}{\delta},\frac{1}{\delta})}\biggr].
\end{eqnarray}

 \par The Laplace transform of this
density also provides the moment generating function of the extended
gamma density thereby the moment generating functions of extended
form of Weibull, chisquare, Reyleigh, Maxwell-Boltzmann, exponential
and other densities in this general class. Usually we do not find
the moment generating function or Laplace transform and the
characteristic function of the generalized gamma density in the
literature when $\delta\neq1$. Here we obtained the Laplace
transform of the generalized gamma density, besides giving an
extension to this density.
\vskip.3cm
\noindent
\emph{1.5 \hskip.3cm Multivariate
generalizations}
\vskip.3cm
 One generalization  of the model in (\ref{pathwayless1}) for one scalar case is give by
\begin{eqnarray}\label{multipathwayless1}
f_{1}(x_1,x_2,\cdots,x_n)&=&K x_1^{\gamma_1-1}
x_2^{\gamma_2-1} \cdots x_n^{\gamma_n-1}\nonumber\\&& \times
[1-(1-\alpha)(a_1x_1^{\delta_1}+a_2x_2^{\delta_2}+\cdots+a_nx_n^{\delta_n})]^{{\frac{\eta}
{1-\alpha}}},
\end{eqnarray}
$\alpha<1,~\eta>0,~a_i>0,~\delta_j>0,~i=1,~2,\cdots,n,~1-(1-\alpha)\sum_{i=1}^{n}a_i x_i ^{\delta_i}>0.$
We can see that (\ref{multipathwayless1}) is the Dirichlet family of densities. For $\alpha<1$, (\ref{multipathwayless1}) stays in the type-1 Dirichlet form and for $\alpha>1$ it stays as a type-2 Dirichlet form. This multivariate analogue can also produce multivariate extensions
to Tsallis statistics and superstatistics. Here the variables are
not independently distributed, but when $\alpha\rightarrow 1$ we
have a surprising result that $x_1,~x_2,\cdots,x_n$ will become
independently distributed generalized gamma variables. Various generalizations of the pathway model are considered by Mathai and associates. The normalizing constants can be obtained
by integrating out the variables one at a time, starting from $x_n$ an going to $x_1$ (see Mathai \& Provost (2006)).
\vskip.3cm
\noindent
{\bf 2 \hskip.3cm Connections to astrophysics and statistical mechanics}
\vskip.3cm
\noindent
\emph{2.1 \hskip.3cm Superstatistics consideration and pathway model }
\vskip.3cm
Beck and Cohen (2003) developed the concept of superstatistics
in statistical mechanics. From a statistical point of view, the
procedure is equivalent to starting with a conditional density for a
random variable $x$ at a given value of a parameter $\theta$. Then
assume the parameter $\theta$ has a prior density. Consider the
conditional density of the form
\begin{equation}\label{fx|theta}
f_{x|\theta}(x|\theta)=k_1 x^{\gamma}{\rm e}^{-\theta
x^\delta},~\gamma+1>0,~\theta>0,~\delta>0,~x>0,
\end{equation}
where $k_1=\frac{\delta
\theta^{\frac{\gamma+1}{\delta}}}{\Gamma(\frac{\gamma+1}{\delta})}$.
Suppose that $\theta$ has an exponential density given by
$f_{\theta}(\theta)=\lambda {\rm e}^{-\lambda
\theta},\\~\lambda>0,~\theta>0.$ Then the unconditional density of $x$
is given by
\begin{eqnarray}\label{fxx}
f_{x}(x)&=&\int_{\theta}f_{x|\theta}(x|\theta)f_{\theta}(\theta)
{\rm d}\theta\nonumber\\
&=&\frac{\delta\Gamma(\frac{\gamma+1}{\delta}+1)x^\gamma
[1+\frac{x^\delta}{\lambda}]^{-({\frac{\gamma+1}{\delta}+1})}}
{\lambda^\frac{\gamma+1}{\delta}\Gamma(\frac{\gamma+1}{\delta})},
\end{eqnarray}
(see Beck \& Cohen (2003), Beck (2006), Mathai \& Haubold (2007a), Mathai et al. (2010)). Equation (\ref{fxx}) is the superstatistics of Beck \& Cohen (2003), in the sense of superimposing
another distribution or the distribution of $x$ with superimposed
distribution of the parameter $\theta.$ In a physical problem the
parameter $\theta$ may be something like temperature having its own
distribution. Several physical interpretations of superstatistics
are available from the papers of Beck and others. The factor
$[1+\frac{x^\delta}{\lambda}]^{-({\frac{\gamma+1}{\delta}+1})}$
written as $ [1+(\alpha-1)x^\delta]^{-\frac{1}{\alpha-1}},~\alpha>1$ is the
foundation for the current hot topic of Tsallis statistics in
non-extensive statistical mechanics. Observe that only a form of the type $[1+\frac{x^\delta}{\lambda}]^{-({\frac{\gamma+1}{\delta}+1})}$ where $1+\frac{x^\delta }{\lambda}>0,~\lambda>0,~x^\delta
>0,~{\frac{\gamma+1}{\delta}+1}>0$, that is, only a type-2 beta form can come from such
a consideration. In other words a type-1 beta form cannot come
because for the convergence of the integral in (\ref{fxx}), $a+x^\delta $ must be
positive with $\lambda>0$ and $x^\delta >0$. It is to be pointed out here that the superstatistics of Beck
and Cohen (2003) and Beck (2006) are available from the procedure given above. It goes without saying that
only type-2 beta form as given in (\ref{pathwaygreater1})
is available from superstatistics considerations, see Seema Nair \& Kattuveettil (2010).
The conditional density of the
random variable $\theta$, given $x$ is the posterior probability
density of $\theta$ and is given by
\begin{eqnarray}\label{ftheta|x}
f_{\theta|x}(\theta|x)&=&\frac{f_{\theta}(\theta)
f_{x|\theta}(x|\theta)}{f_{x}(x)}\nonumber\\&=&
\frac{\lambda^{{\frac{\gamma+1}{\delta}}+1}[1+\frac{x^\delta}{\lambda}]
^{{\frac{\gamma+1}{\delta}}+1}}
{\Gamma(\frac{\gamma+1}{\delta}+1)}{\rm
e}^{-\theta(\lambda+x^\delta)}\theta^{\frac{\gamma+1}{\delta}},~\theta>0.
\end{eqnarray}
With the help of (\ref{ftheta|x}) we can obtain the Bayes' estimate of the
parameter $\theta$. To this extent, let
\begin{eqnarray}\label{phix}
\Phi(x)&=&E_{\theta|x}(\theta|x)=\int_0^{\infty}\theta
f_{\theta|x}(\theta|x){\rm d}\theta
\nonumber\\&=&\frac{\frac{\gamma+1} {\delta}+1}{\lambda+x^\delta} .
\end{eqnarray}
Superstatistics and Tsallis statistics in statistical mechanics are given interpretations in terms of
Bayesian statistical analysis. Subsequently superstatistics is extended by replacing each component of the
conditional and marginal densities by Mathai's pathway model and further both components are replaced by
Mathai's pathway model. This produces a wide class of mathematically and statistically interesting functions
for prospective applications in statistical physics (Mathai \& Haubold (2010)). The same procedure can be used to look at the extended forms. Let
the conditional density of $x$, given $\theta$, be of the form
\begin{equation}\label{falphax|theta}
f_{\alpha_{x|\theta}}(x|\theta)=k_2x^{\gamma}
[1+\theta(\alpha-1)x^\delta]^{-{\frac{1}
{\alpha-1}}},~x>0,~\theta>0,~\alpha>1,~\delta>0,
\end{equation}
where $k_2=\frac{\delta(\theta(\alpha-1))^{\frac{\gamma+1}{\delta}}
\Gamma(\frac{1}{\alpha-1})}{\Gamma(\frac{\gamma+1}{\delta})
\Gamma(\frac{1}{\alpha-1}-\frac{\gamma+1}{\delta})}$ and assume that
the parameter $\theta$ has a prior density
$f_{\theta}(\theta)=\lambda {\rm e}^{-\lambda
\theta},~\lambda>0,~\theta>0.$ Then the unconditional density of $x$
is given by
\begin{equation}\label{falphax}
f_{\alpha_{x}}(x)=\frac{\lambda\delta
x^{-(\delta+1)}}{(\alpha-1)\Gamma(\frac{\gamma+1}{\delta})
\Gamma(\frac{1}{\alpha-1}-\frac{\gamma+1}{\delta})}
G_{1,2}^{2,1}\biggl[\frac{\lambda}{x^{\delta} (\alpha-1)}
 \bigg|_{0,\frac{1}{\alpha-1}-\frac{\gamma+1}{\delta}
-1}^{-\frac{\gamma+1}{\delta}}\biggr],~x>0,
\end{equation}
where $G_{1,2}^{2,1}(\cdot)$ is a $G-$function. For the theory and application of the functions see Mathai (1993).
 The posterior
probability density is given by
\begin{equation}\label{falphatheta|x}
f_{\alpha_{\theta|x}}(\theta|x)=c_4^{-1}(\alpha-1)^{\frac{\gamma+1}{\delta}+1}
x^{\gamma+\delta+1}\Gamma(\frac{1}{\alpha-1})\theta^{\frac{\gamma+1}{\delta}}{\rm
e}^{-\lambda \theta}[1+\theta(\alpha-1)x^\delta]^{-{\frac{1}
{\alpha-1}}},
\end{equation}
where
$c_4=G_{1,2}^{2,1}\biggl [\frac{\lambda}{x^{\delta} (\alpha-1)}
 \bigg|_{0,\frac{1}{\alpha-1}-\frac{\gamma+1}{\delta}
-1}^{-\frac{\gamma+1}{\delta}}\biggr].$
 Then the Bayes' estimate of $\theta$, at given $x$, defined by
$\Phi_{\alpha}(x)$, is given by the following:
\begin{equation}\label{phialphax}
\Phi_{\alpha}(x)=E_{\alpha_{\theta|x}}(\theta|x)=\frac{c_4^{-1}}{(\alpha-1)
x^\delta}G_{1,2}^{2,1}\biggl [\frac{\lambda}{x^{\delta} (\alpha-1)}
 \bigg|_{0,\frac{1}{\alpha-1}-\frac{\gamma+1}{\delta}
-2}^{-1-\frac{\gamma+1}{\delta}}\biggr].
\end{equation}

\noindent{\sc{Lemma 1.}}{\it{
Let the conditional density of $x$ given $\theta$ be
$f_{x|\theta}(x|\theta)$ as in (\ref{fx|theta}) and assume that the parameter
$\theta$ has a prior density $f_{\theta}(\theta)=\lambda {\rm
e}^{-\lambda \theta},~\lambda>0,~\theta>0.$ Then the Bayes' estimate
of $\theta$ is given by $\Phi(x)$ in (\ref{phix}).}}\\\\
\noindent{\sc{Theorem 3.}}{\it{
Let the conditional density of $x$ given $\theta$ be
$f_{\alpha_{x|\theta}}(x|\theta)$ as in (\ref{falphax|theta}) and assume that the
parameter $\theta$ has a prior density $f_{\theta}(\theta)=\lambda
{\rm e}^{-\lambda \theta},~\lambda>0,~\theta>0.$ Then the Bayes'
estimate of $\theta$ is given by $\Phi_{\alpha}(x)$ in (\ref{phialphax}). }}\\
\noindent
Thus, the popular superstatistics in statistical mechanics can be considered as a special case of the pathway model in (\ref{fxx}) for $\eta=1$ and $\delta=1$.
\vskip.3cm
\noindent
\emph{2.1.1 \hskip.3cm$\alpha$-gamma models associated with Bessel function}
\vskip.3cm
Sebastian (2011) deals with a new family of statistical
  distributions associated with Bessel function which gives an extension of
  the gamma density, which will connect the fractional calculus and statistical
  distribution theory through the special function. The idea is motivated by the fact that a non-central chi-square
density is associated with a Bessel function.
 In order to make thicker or thinner tails
 in a gamma density we consider a density function of the following type:
 \begin{equation}\label{gener2}
f_{x|a}(x|a)=\frac{\rho ~a^{\frac{\gamma}{\rho}}{\rm e}
^{-\frac{\delta}{a}}}{\Gamma(\frac{\gamma}{\rho})} x^{\gamma-1}{\rm
e}^{-ax^{\rho}}{_0F_1}(~;\frac{\gamma} {\rho};\delta x^\rho);~0<
x<\infty
\end{equation}
Where ${_0F_1}(~;b;
x)=\sum_{k=0}^\infty \frac{{( x)}^k}{(b)_k~k!},~(b)_k
$ is the Pochhammer symbol,
$(b)_m=b(b+1)\cdots(b+m-1),b\neq0,(b)_0=1.$ When $\delta=0$
the equation (\ref{gener2}) reduces to generalized gamma density. Note that
this is the generalization of some standard statistical densities
such as gamma, Weibull, exponential, Maxwell-Boltzmann, Rayleigh and
many more. When $\delta=0,\rho=2,$ (\ref{gener2}) reduces to folded standard
normal density. We can extend the generalized gamma model associated with Bessel
function in (\ref{gener2})  by using the pathway model of Mathai (2005), when
$\alpha<1$ we get the extended function as

\begin{eqnarray}
g_{\alpha}(x)&=&k_1
x^{\gamma-1}[1-a(1-\alpha)x^\rho]^{\frac{1}{1-\alpha}}
{_0F_1}(~;\frac{\gamma} {\rho};\delta
x^\rho);~\alpha<1,\nonumber\\
&&a>0,\rho>0,1-a(1-\alpha)x^\rho>0, x>0,
\end{eqnarray}
where $k_1$ is the normalizing constant.  Note that $g_{\alpha}(x)$
is a generalized type-1 beta model associated with Bessel function. 
 Observe that for $\alpha>1,$ writing
$1-\alpha=-(\alpha-1)$ in equation (\ref{gener2}) produces extended type-2
beta form which is given by
\begin{equation}
f_{\alpha}(x)=k_2
x^{\gamma-1}[1+a(\alpha-1)x^\rho]^{-\frac{1}{\alpha-1}}
{_0F_1}(~;\frac{\gamma} {\rho};\delta x^\rho);~\alpha>1,a>0,\rho>0,
\end{equation}
where $k_2$ is the normalizing constant. Note that in both the
cases, when $\alpha\rightarrow1$, we have (\ref{gener2}) and hence it can be
considered to be an extended form of (\ref{gener2}). This model has wide potential applications in the discipline
 physical science especially in statistical mechanics, see  Sebastian (2009, 2011).
 \begin{figure}
\begin{center}
\resizebox{7cm}{5cm}{\includegraphics{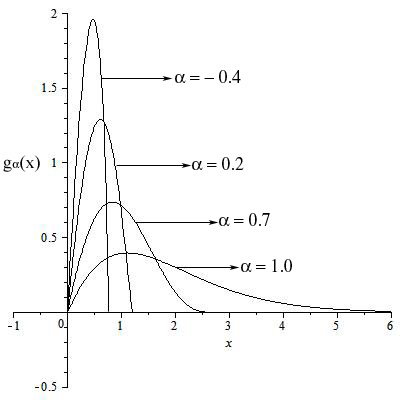}}
\resizebox{7cm}{5cm}{\includegraphics{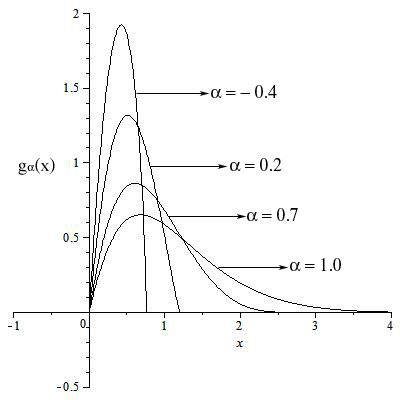}}\\
\caption {\small (a)
$\alpha$ gamma bessel model for $\delta=0.5,\alpha<1$
\hskip1cm (b) $\alpha$ gamma bessel model for $\delta=-0.5,\alpha<1$\label{ppplot2}}
\end{center}
\end{figure}
 \begin{figure}
\begin{center}
\resizebox{7cm}{5cm}{\includegraphics{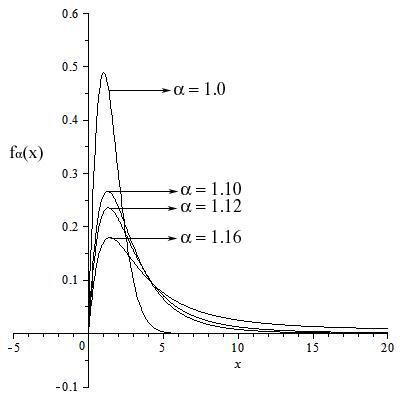}}
\resizebox{7cm}{5cm}{\includegraphics{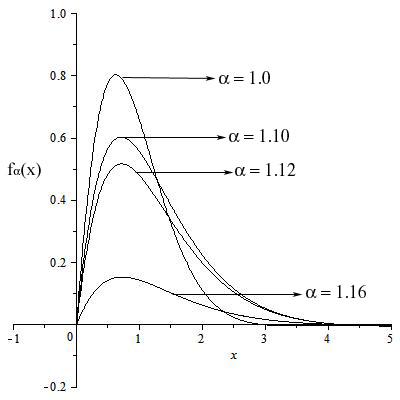}}\\
\caption {\small (a)
$\alpha$ gamma bessel model for $\delta=0.5,\alpha>1$
\hskip1cm (b) $\alpha$ gamma bessel model for $\delta=-0.5,\alpha>1$\label{ppplot3}}
\end{center}
\end{figure}

For fixed values of $\gamma=2, \rho=1.2$ and $a=1$, we can look at the graphs for
$\delta=-0.5, q<1$, $\delta=0.5, q<1$ as well as for $\delta=-0.5, q>1$, $\delta=0.5, q>1$. From the Figures \ref{ppplot2}, we can see that when $q$ moves from $-\infty$
to 1,
 the curve becomes less peaked.
Similarly from Figure \ref{ppplot3},  we can see that when $q$ moves from 1
to $\infty$, the curve becomes less peaked.
It is also observed that
when $\delta>0$
the right tail of the density becomes
thicker and thicker. Similarly when $\delta<0$ the right tail gets thinner
and thinner. Densities exhibiting thicker or thinner tail occur frequently in many different
areas of science. For practical
purposes of analyzing data from physical experiments and in building up models in statistics,
we frequently select a member from a parametric family of distributions. But it is often found
that the model requires a distribution with a thicker or thinner tail than the ones available
from the parametric family.

%

\vskip.3cm
\noindent
\emph{2.2 \hskip.3cm Tsallis statistics}
\vskip.3cm
Model (\ref{pathwayless1}) for $\gamma=1, \delta=1, a=1, \eta=1$ is Tsallis
statistics, which is the foundation for the newly created hot topic
of non-extensive statistical mechanics. It is stated that over three
thousand papers are written on Tsallis statistics so far. With
$\alpha$ replaced in $-\alpha$ in Model (\ref{pathwayless1}), with
$\gamma=0, a=1,\eta=1,\delta=1,$ one has an extension of the
exponential function, known as $q$-exponential function [The
parameter $q$ is used instead of $\alpha$ and hence
$q$-exponential]. The basis for the current hot topic of
\emph{$q$-calculus} is this $q$-exponential function. When
$\gamma=1, a=1,\eta=1,\delta=1$ one has the following property if
the resulting function of $f(x)$ is divided by the resulting
normalizing factor $c$ and obtain the function $g(x)$, that is,
$g(x)=\frac{f(x)}{c}$. Then
\begin{equation*}
  \frac{  {\rm d}g(x)}{{\rm d}x}=-[g(x)]^\alpha.
\end{equation*}This is the power law in physics literature. Thus, power function law is a special case of the pathway model of (\ref{pathwayless1}) and (\ref{pathwaygreater1}) for $\gamma=1,~\delta=1,~a=1,~\eta=1$ and exactly the normalizing constants $c_1$ and $c_2$. Also Tsallis statistics can be looked upon as a special case of (\ref{pathwayless1}) and (\ref{pathwaygreater1}) for $\gamma=1,~\delta=1,~a=1,~\eta=1$.
\vskip.3cm
\noindent
\emph{2.3 \hskip.3cm Extension of thermonuclear functions through pathway model}
\vskip.3cm
Nuclear reactions govern major aspects of the chemical evolution of
the universe. A proper understanding of the nuclear reactions that
are going on in hot cosmic plasmas, and those in the laboratories as
well, requires a sound theory of nuclear-reaction dynamics. The
reaction probability integral is the probability per unit time that
two particles, confined to a unit volume, will react with each other. Practically all applications of fusion plasmas are controlled in some way or other by the theory of thermonuclear reaction rates under specific circumstances. After several decades of effort, a systematic and complete theory of thermonuclear reaction rates has been developed (Haubold \& John (1978); Anderson et al. (1994); Mathai \& Haubold (1988); Mathai \& Haubold (2002)).
\par The standard thermonuclear function in the Maxwell-Boltzmann case in the theory of nuclear reactions, is given by the following (Haubold \& Mathai (1985); Mathai \& Haubold (1988)):\\

\noindent
{\it Nonresonant case with depleted tail}:
\begin{equation}\label{I1}
I_{1}=\int_{0}^{\infty}x^{\gamma-1}e^{-ax^{\delta}-bx^{-\rho}}{\rm
d}x,~~a>0,~b>0,~\delta>0,~\rho>0.
\end{equation}
{\it Nonresonant case with depleted tail and high energy cut-off}:
\begin{equation}\label{I2}
I_2=\int_{0}^{d}x^{\gamma-1}e^{-ax^{\delta}-bx^{-\rho}}{\rm
d}x,~~a>0,~b>0,~\delta>0,~\rho>0,~d<\infty.
\end{equation}
Note that if $\delta=1$ is taken as the standard Maxwell-Boltzmannian behavior, then for $\delta >1$
the right tail will deplete faster and if $\delta <1$ then the depletion will be slower in (\ref{I2}). We can
extend the thermonuclear functions given in (\ref{I1}) and (\ref{I2}) to more general classes by replacing ${\rm e}^{-b x^{-\rho}}$ by a binomial factor $[1-b(1-\alpha)x^{-\rho}]^{\frac{1}{1-\alpha}}$ (Mathai \& Haubold (2008), Joseph \& Haubold (2010), Haubold \& Kumar (2008)).
Thus we consider the general class of reaction rate integrals
\begin{equation}\label{I1alpha}
I_{1\alpha}=\int_{0}^{\infty}x^{\gamma-1}{\rm
e}^{-ax^{\delta}}[1+b(\alpha-1)x^{-\rho}]^{-{\frac{1}
{\alpha-1}}}{\rm d}x,~a>0,~b>0,~\delta>0,~\rho>0,~\alpha>1,
\end{equation}
\begin{equation}\label{I2alpha}
I_{2\alpha}=\int_{0}^{d}x^{\gamma-1}{\rm e}^{-ax^{\delta}}[1-b(1-\alpha)x^{-\rho}]^{\frac{1}{1-\alpha}}
{\rm d}x,~a>0,~b>0,~\delta>0,~\rho>0,~\alpha<1.
\end{equation}
We can evaluate these extended integrals by using Mellin convolution property.
(\ref{I1alpha}) can be looked upon as the Mellin convolution of two  real positive scalar independently distributed random variables $x_1$ and
$x_2$, where $x_1$ has a
generalized gamma density and  $x_2$ has an extended form of stretched exponential function.
Make the transformation $u=x_1x_2$ and $v=x_1$, and then proceed as in the case of the evaluation of the Laplace transform of the pathway model, we can arrive at the value of the extended reaction rate integral as
\begin{eqnarray}\label{I1alphahfunction}
I_{1 \alpha}&=&\int_{0}^{\infty}{\rm
e}^{-ax^{\delta}}x^{\gamma-1}[1+b(\alpha-1)x^{-\rho}]^{-{\frac{1}
{\alpha-1}}}{\rm d}x\nonumber\\&=&\frac{1}{\rho \mu a^
{\frac{\gamma}{\delta}}\Gamma(\frac{1}{\alpha-1})}
H_{1,2}^{2,1}\biggl[a^{\frac{1}{\delta}} (b(\alpha-1))^
{\frac{1}{\rho}}
 \bigg|_{(0,\frac{1}{\rho}),(\frac{\gamma}{\delta},{\frac{1}{\delta}})}^
 {(1-\frac{1}{\alpha-1},\frac{1}{\rho})}
\biggr],~b=u^{\rho}.
\end{eqnarray}
Similarly we can evaluate (\ref{I2alpha}) by considering it as the Mellin convolution of exponential
density and pathway model for $\alpha<1$. When we take the limit as $\alpha\rightarrow1,$ in (\ref{I1alphahfunction}), we will get the value of the reaction rate integral given in (\ref{I1}) and (\ref{I2}), as an $H$- function.
\vskip.3cm
\noindent
\emph{2.3.1 \hskip.3cm Inverse Gaussian as a particular case of the pathway model}
\vskip.3cm
\noindent Note that one form of the inverse Gaussian probability
density function is given by
\begin{equation*} g(x)=kx^{-\frac{3}{2}}{\rm
e}^{-\frac{\lambda}{2}(\frac{x}{\mu^2}+\frac{1}{x})},~\mu\neq0,~x>0,~\lambda>0,
\end{equation*}
where $k$ is the normalizing constant (see Mathai (1993)). Put
 $\gamma=-\frac{5}{2},~\delta=1,~\rho=1,~a=\frac{\lambda}{2 \mu^2},~b=\frac{\lambda}{2}$
 in equation (\ref{I1}),  we can see that the inverse Gaussian density is the
 integrand in $I_1$. Hence $I_1$ in equation (\ref{I1}) can be used to evaluate the moments
 of inverse Gaussian density. Hence the integrand in the extended integral $I_{1\alpha}$ can be considered as the
extended form of the inverse Gaussian density.
\vskip.3cm
\noindent
\emph{2.3.2 \hskip.3cm An interpretation of the pathway parameter $\alpha$}
\vskip.3cm
Let us start with the pathway density in (\ref{pathwayless1}) with $\eta=1$. For this to remain a density we need the condition $1-a(1-\alpha)x^\delta>0$ or when $x$ is positive then $0<x<\frac{1}{[a(1-\alpha)]^{\frac{1}{\delta}}},\alpha<1,$ or if we have the model in (\ref{I2alpha}) then
$$-\frac{1}{[a(1-\alpha)]^{\frac{1}{\delta}}}<x<\frac{1}{[a(1-\alpha)]^{\frac{1}{\delta}}},~\alpha<1.$$
Outside this range, the density is zero. Thus for $x>0$ the right tail of the density is cut-off at $\frac{1}{[a(1-\alpha)]^{\frac{1}{\delta}}}$. If $d$ is this cut-off on the right then

\begin{equation}\label{x}
d=\frac{1}{[a(1-\alpha)]^{\frac{1}{\delta}}}\Rightarrow\alpha=1-\frac{1}{ad^\delta},~ \text{for}~\alpha<1.
\end{equation}
As $\alpha$ moves closer to 1 then the cut-off point $d$ moves farther out and eventually when $\alpha\rightarrow1$ then $d\rightarrow\infty$. In this case $\alpha$ is computed easily from the cut-off. Also the pathway parameter $\alpha$ can be estimated in terms of arbitrary moments, see for example Mathai \& Haubold (2007a).
\vskip.3cm
\noindent
{\bf 3 \hskip.3cm Pathway model and fractional calculus}
\vskip.3cm
When solving certain problems in reaction-diffusion, relaxation-oscillations, it is observed that the solution is obtained in terms of exponential function, see Haubold and mathai (2000, 1995). But if a fractional integration is under consideration then the residual reaction equation is given by
\begin{equation}\label{fractional}
N(t)=N_0-c^\nu {_0D_t}^{-\nu}N(t),~~~\nu>0
\end{equation}
where ${_0D_t}^{-\nu}$ is the standard Riemann-Louville fractional integral operator, where $N(t)$ is the number density of the reacting particles and (\ref{fractional}) has the solution in terms of generalized Mittag-Leffler function. While fitting such a model Mittag-leffler function will provide a better fit as compared to exponential function. If we consider the residual rate of change $c$, in (\ref{fractional}) is a random variable having a gamma type density

\begin{equation}\label{gamma1}
g(c)=\frac{w^\mu}{\Gamma(\mu)}c^{\mu-1}{\rm e}^{-w c},~~w>0,~0<c<\infty
\end{equation}
where $\mu>0,~w,~\mu$ are known and $\frac{\mu}{w}$ is the mean value of $c$. The residual rate of change may have small probabilities of it being too large or too small and the maximum probability may be for a medium range of values for the residual rate of change $c$. (\ref{fractional}) is the situation where the residual rate of change is such that the production rate dominates so that we have the form $-c^\nu, \nu>0,c>0$. If the destruction rate dominates then the constant will be of the form $c^\nu,~~\nu>0,~c>0$. By solving (\ref{fractional}), we have the unconditional number density of the following form:

\begin{equation}
N(t)=\frac{N_0}{\Gamma(\mu)}t^{\mu-1}(1+\frac{t^\nu}{w^\nu})^{-(\gamma+1)},~~~~0<t<\infty,~w>0.
\end{equation}

If we make the substitution $\gamma+1=\frac{1}{\alpha-1}$, $\alpha>1\Rightarrow\gamma=\frac{\alpha-2}{\alpha-1}$ and $w^{-\nu}=b(\alpha-1),~b>0$. Then we have
\begin{equation}\label{Number}
N(t)=\frac{N_0}{\Gamma(\mu)}t^{\mu-1}[1+b(\alpha-1)t^\nu]^{-\frac{1}{\alpha-1}}
\end{equation}
for $\alpha>1,~t>0,~b>0,~\mu>0$. For general values of $\mu$ and $\alpha>1$ such that $\frac{1}{\alpha-1}-\frac{\mu}{\nu}>0$, (\ref{Number}) corresponds to the pathway model of Mathai as well as the superstatistics of Beck and Cohen (2003). As an application of pathway model in fractional calculus, a general pathway fractional integral operator is introduced in Seema Nair (2009), which generalizes the classical Riemann-Liouville fractional integration operator, see section 4.5.
\vskip.3cm
\noindent
\emph{3.1 \hskip.3cm $\mathcal {P}$-transform }
\vskip.3cm

Consider the generalized Kr\"{a}tzel function $D_{\rho,\beta}^{\nu,\alpha }(x)$, dealt with in Kilbas and Kumar (2009), given by
  \begin{equation}\label{ptransformkernal}
D_{\rho,\beta}^{\nu,\alpha}(x)=\int _{ 0 }^{\infty} y^{\nu-1}[1+a(\alpha-1
)y^\rho]^{-\frac{1}{\alpha-1 }} {\rm e}^{-xy^{-\beta }}{\rm d}y,~ x>0,
\end{equation}
with $ \gamma \in {\mathbb C},\beta>0, \alpha>1$.  The generalized Kr\"{a}tzel function is obtained by using the pathway model introduced by Mathai (2005). The $\mathcal {P}$-transform or pathway transform introduced in
Kumar (2009) by using the pathway idea is defined as
\begin{equation}\label{ptransform1}
({\mathcal P}_{\nu}^{\rho,\beta,\alpha} f)(x)=\int_0^\infty
D_{\rho,\beta}^{\nu,\alpha }(xt)f(t){\rm d}t, x>0, \end{equation}
where $D_{\rho,\beta}^{\nu,\alpha }(x)$ denotes the function given
in (\ref{ptransformkernal}). The ${\mathcal P}$-transform is defined in the space $\mathbb{L}_{\nu,r}$ consisting of the lebesgue
 measurable complex valued functions $f$ for which
\begin{equation}\label{ptransform2}
\| f \|_{\nu,r}=\left\{\int_0^\infty |t^\nu f(t)|^r \frac{{\rm
d}t}{t}\right\}^{\frac{1}{r}}<\infty, \end{equation} for $1 \leq r <
\infty, \nu \in \mathbb{R}$. When $\beta=1, a=1$ and
$\alpha\rightarrow1$ the $\mathcal{P}$-transform reduces to the
Kr\"{a}tzel tranform, introduced by  Kr\"{a}tzel (1979), which is
given by
\begin{equation}\label{ptransform3}{\mathcal K}_\nu^{(\rho)}f(x)= \int_0^\infty
Z_\rho^\nu(xt)f(t){\rm d}t,~x>0,\end{equation}
where
\begin{equation}\label{ptransform4}
Z_\rho^\nu(x)=\int _{ 0 }^{\infty} y^{\nu -1} {\rm
e}^{-y^\rho-xy^{-1 }}{\rm d}y.\end{equation}
The $\mathcal {P}$-transform reduces to the Meijer transform when
$\rho=1$ and $\alpha \rightarrow 1$. When $a=1, \beta=1$ and
$\alpha\rightarrow1$, then the generalized Kr\"{a}tzel function
defined in (\ref{ptransformkernal}) reduces to the modified Bessel function of the
third kind or Mc Donald function (Erd\'{e}lyi et al. (1953)). Kilbas \& Kumar (2009) have considered  (\ref{ptransformkernal}) for $\beta=1$
and established its composition with fractional operators and
represented it in terms of various generalized special functions.
The Kr\"{a}tzel function defined in (\ref{ptransform4}) for any real $\rho$, was
studied by Kilbas et al. (2006) and established its representations
in terms of $H$-function and extended the function from positive
$x>0$ to complex $z$. Here we establishes connection between
generalized Kr\"{a}tzel function and $\mathcal {P}$-transform with
generalized special functions. \\
\begin{figure}
\begin{center}
~~~~~ \resizebox{6cm}{6cm}{\includegraphics{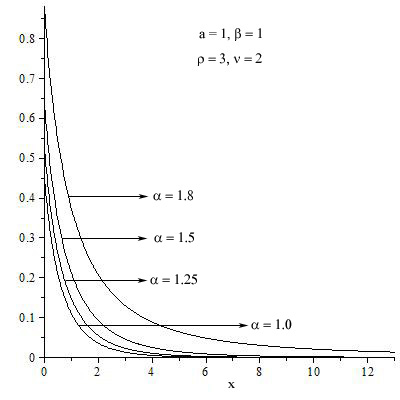}}
~~~~ \resizebox{6cm}{6cm}{\includegraphics{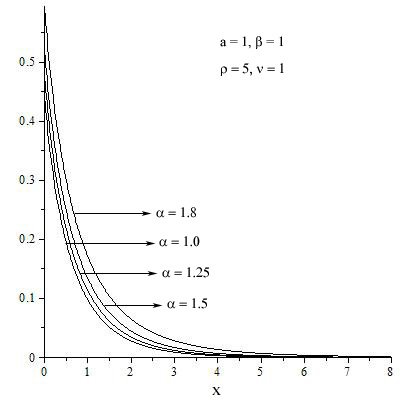}}\\
\caption {\small (a)
Behaviour of $D_{3,1}^{2,\alpha}(x)$ for various values of $\alpha>1$ and $\alpha\rightarrow1$
\hskip1.5cm (b) Behaviour of $D_{5,1}^{2,\alpha}(x)$ for various values of $\alpha>1$ and $\alpha\rightarrow1$\label{ppplot4}}
\end{center}
\end{figure}
The particular case of the kernel  of the $\mathcal{P}$-transform
given in (\ref{ptransformkernal}) is the extended non-resonant thermonuclear function
used in Astrophysics which is already discussed.
As $\alpha\rightarrow 1$ we get the standard reaction rate
probability integral in the Maxwell-Boltzmann case. The behavior of the generalized Kr\"{a}tzel function (\ref{ptransformkernal}) can be
studied from the following graphs.  We take $a=1, \beta=1, \nu=2,
\rho=3$ and $a=1, \beta=1, \nu=2,
\rho=5$ for example and  investigate
the behavior of $D_{3,1}^{2,\alpha}(x)$ and $D_{5,1}^{2,\alpha}(x)$
for various values of $\alpha>1$ and $\alpha\rightarrow1$.  The
graphs of these functions $D_{3,1}^{2,\alpha}(x)$ and
$D_{5,1}^{2,\alpha}(x)$ at
$\alpha=1,\alpha=1.25,~\alpha=1.5,~\alpha=1.8$ are given in Figure
\ref{ppplot4}(a) and \ref{ppplot4}(b), respectively.  From these graphs we observe that if
the value of the parameter $\alpha$ increases then the curves move
away from the limiting situation for $\alpha\rightarrow1$.\\\\

\noindent
{\bf 4 \hskip.3cm A matrix variate pathway model}
\vskip.3cm
Let $X=(x_{ij}),~i=1,\dots,p,~j=1,\dots,q,~q \geq p ,$ of rank $p$ and of real scalar
variables $x_{ij}'s$ for all $i$ and $j$, subject to the condition that the rank of $X$ is $p$, having the density $f(X)$, where $f(X)$ is a scalar function of $X$ is given by
\begin{equation}\label{matrixpathway}
f(X)=C|A^{\frac{1}{2}}(X-M)B(X-M)'A^{\frac{1}{2}}|^{\gamma}|I-a(1-\alpha)A^{\frac{1}{2}}(X-M)B(X-M)'A^{\frac{1}{2}}|^{\frac{\eta}{1-\alpha}}
\end{equation}
where $C$ is the normalizing constant, $A=A'>0, B=B'>0, A$ is
$p\times p, B$ is $q\times q$, $A$ and $B$ are constant positive
definite matrices, and $M$ is a $p\times q$ constant matrix.
$A^{\frac{1}{2}}$ denotes the positive definite square root of $A$,
$a>0$ and $\alpha$ is the pathway parameter. For keeping
non-negativity of the determinant in (\ref{matrixpathway}) we need the condition
\begin{equation*}
I-a(1-\alpha)A^{\frac{1}{2}}(X-M)B(X-M)'A^{\frac{1}{2}}>0
\end{equation*}
where, for example, the notation $A<Y<B\Rightarrow A=A'>0, B=B'>0,
Y=Y'>0, Y-A>0, B-Y>0$. In (\ref{matrixpathway}) the constant matrix $M$ can act as a
relocation matrix or as the mean value matrix so that $E(X)=M$ where
$E$ denotes the expected value.
\vskip.3cm
\noindent
\emph{4.1 \hskip.3cm The normalizing constants}
\vskip.3cm
This requires certain transformations and the knowledge of the
corresponding Jacobians. Here we will use a few
multi-linear and some nonlinear transformations and the
corresponding Jacobians. The details of the derivations of
these Jacobians are available from
Mathai (1997). Put
\begin{equation*}
Y=A^{\frac{1}{2}}(X-M)B^{\frac{1}{2}}\Rightarrow {\rm d}Y=|A|^{\frac{q}{2}}|B|^{\frac{p}{2}}{\rm d}X,
\end{equation*}
where $|(\cdot)|$ denotes the determinant of $(\cdot)$, ${\rm d}X$ is defined as the following wedge product of differentials:
 \[{\rm d}X={\rm d}x_{11}\wedge{\rm d}x_{12}\wedge\cdots\wedge{\rm d}x_{1q}\wedge{\rm d}x_{22}\wedge\cdots\wedge{\rm d}x_{2q}\wedge\cdots\wedge{\rm d}x_{pq}.\]
see Mathai (1997) for the Jacobian. Since $f(X)$ is assumed to be a density,
\begin{equation*}
1=\int_Xf(X){\rm d}X=C|A|^{-\frac{q}{2}}|B|^{-\frac{p}{2}}\int_Y|YY'|^{\gamma}|I-a(1-\alpha)YY'|^{\frac{\eta}{1-\alpha}}{\rm d}Y.
\end{equation*}
Now, make suitable transformations and integrating out
over the Stiefel manifold (for details, see Mathai (2005)) we have
\begin{align}\label{1}
1&=\int_Xf(X){\rm d}X\notag\\
&=C|A|^{-\frac{q}{2}}|B|^{-\frac{p}{2}}\frac{\pi^{\frac{pq}{2}}}{\Gamma_p(\frac{q}{2})}\int_Z|Z|^{\gamma+\frac{q}{2}-\frac{p+1}{2}}|I-a(1-\alpha)Z|^{\frac{\eta}{1-\alpha}}{\rm d}Z
\end{align}
where, for example,
\begin{equation}
\Gamma_p(\beta)=\pi^{\frac{p(p-1}{4}}\Gamma(\beta)\Gamma(\beta-\frac{1}{2})...\Gamma(\beta-\frac{p-1}{2}), \Re(\beta)>\frac{p-1}{2}
\end{equation}
is the real matrix-variate gamma and $\Re(\cdot)$ denotes the real
part of $(\cdot)$.\\

 Further evaluation of
the integral in (\ref{1}) depends on the value of $\alpha$. If $\alpha
<1$ then (\ref{1}) belongs to the real matrix-variate type-1 beta. Then
integrating out with the help of a matrix variate type-1 beta integral we get, for $q\ge p, \alpha <1$,

\begin{equation}\label{c1}
C=|A|^{\frac{q}{2}}|B|^{\frac{p}{2}}[a(1-\alpha)]^{p(\gamma+\frac{q}{2})}\frac{\Gamma_p(\frac{q}{2})
\Gamma_p(\gamma+\frac{q}{2}+\frac{\eta}{1-\alpha}+\frac{p+1}{2})}
{\pi^{\frac{pq}{2}}\Gamma_p(\gamma+\frac{q}{2})\Gamma_p(\frac{\eta}{1-\alpha}+\frac{p+1}{2})}
\end{equation}
for $\alpha<1, \eta >0, a>0,\Re(\gamma+\frac{q}{2})>\frac{p-1}{2}, q\ge p, p,q=1,2,..., A=A'>0, B=B'>0$. For $\alpha >1$, write $1-\alpha = -(\alpha -1), \alpha >1$ then the integral in (\ref{1}) goes into a type-2 beta form. Then integrating out with the help of a real matrix-variate type-2 beta integral we have for $\alpha >1$,
\begin{equation}\label{c2}
C=\frac{|A|^{\frac{q}{2}}|B|^{\frac{p}{2}}[a(\alpha-1)]^{p(\gamma+\frac{q}{2})}\Gamma_p(\frac{q}{2})
\Gamma_p(\frac{\eta}{\alpha-1})}{\pi^{\frac{pq}{2}}\Gamma_p(\gamma+\frac{q}{2})\Gamma_p(\frac{\eta}
{\alpha-1}-\gamma-\frac{q}{2})}
\end{equation}
for $\alpha>1, \Re(\gamma+\frac{q}{2})>\frac{p-1}{2}, \Re(\frac{\eta}{\alpha-1}-\gamma-\frac{q}{2})>\frac{p-1}{2}, a>0, \eta>0, A=A'>0, B=B'>0$. When $\alpha\rightarrow 1$ we have
\begin{equation*}
\lim_{\alpha\rightarrow
1}|I-a(1-\alpha)Z|^{\frac{\eta}{1-\alpha}}={\rm e}^{-a\eta~{\rm
tr}(Z)}
\end{equation*}
and hence, evaluating with the help of a real matrix-variate gamma integral, we will get $C$ for $\alpha\rightarrow 1$,
\begin{equation}\label{c3}
C=\frac{|A|^{\frac{q}{2}}|B|^{\frac{p}{2}}(a\eta)^{p(\gamma+\frac{q}{2})}\Gamma_p(\frac{q}{2})}{\pi^{\frac{pq}{2}}\Gamma_p(\gamma+\frac{q}{2})}
\end{equation}
for $\Re(\gamma+\frac{q}{2})>\frac{p-1}{2}, a>0,\eta >0$.\\
A model corresponding to (\ref{matrixpathway}) in the complex
domain, along with its properties and connections to various other fields, is studied
in Mathai and Provost (2005).
\vskip.3cm
\noindent
\emph{4.2 \hskip.3cm Density of the volume content}
\vskip.3cm The squared volume of the $p$-parallelotope in the
$q$-space,
\begin{equation}\label{v2}
v^2=|A^{\frac{1}{2}}(X-M)B(X-M)'A^{\frac{1}{2}}|.
\end{equation}
The density of $u=v^2$ in (\ref{v2}) can be evaluated by looking at the $h$-th moment
of $v^2$ for an arbitrary $h$. That is, the expected value of $(v^2)^h$, is given by
\begin{equation*}
E(v^2)^h=\int_X(v^2)^hf_5(X){\rm d}X.
\end{equation*}
This is available from the normalizing constants in (\ref{c1}), (\ref{c2}) and (\ref{c3}) by observing that
the only change is that the parameter $\gamma$ is changed to $\gamma+h$. Hence from (\ref{c1}) for $\alpha <1$,
\begin{equation}
E(v^2)^h=[a(1-\alpha)]^{-ph}\frac{\Gamma_p(\gamma+\frac{q}{2}+h)}{\Gamma_p(\gamma+\frac{q}{2})}\frac{\Gamma_p(\gamma+\frac{q}{2}+\frac{\eta}{1-\alpha}+\frac{p+1}{2})}{\Gamma_p(\gamma+\frac{q}{2}+\frac{\eta}{1-\alpha}+\frac{p+1}{2}+h)}
\end{equation}
for $\alpha<1, \Re(\gamma+\frac{q}{2})>\frac{p-1}{2}, \Re(\gamma+\frac{q}{2}+h)>\frac{p-1}{2}, \eta>0, a>0$.
 Let $u_1=[a(1-\alpha)]^pv^2$. Then
\begin{align}\label{Eu1h}
E(u_1^h)&=\prod_{j=1}^p\frac{\Gamma(\gamma+\frac{q}{2}-\frac{j-1}{2}+h)}{\Gamma(\gamma+\frac{q}{2}-\frac{j-1}{2})}\frac{\Gamma(\gamma+\frac{q}{2}+\frac{\eta}{1-\alpha}+\frac{p+1}{2}-\frac{j-1}{2})}{\Gamma(\gamma+\frac{q}{2}+\frac{\eta}{1-\alpha}+\frac{p+1}{2}-\frac{j-1}{2}+h)}\\
&=E(v_1^h)E(v_2^h)...E(v_p^h)\notag
\end{align}
where $v_j$ is a real scalar type-1 beta random variable with the parameters
$(\gamma+\frac{q}{2}-\frac{j-1}{2},\frac{\eta}{1-\alpha}+\frac{p+1}{2}), j=1,...,p$, and further,
$v_1,...,v_p$ are statistically independently distributed. Hence structurally
\begin{equation}\label{u_1}
u_1=v_1v_2...v_p
\end{equation}
and the density of $u_1$ is that of $v_1...v_p$. This is available
by treating (\ref{Eu1h}) as coming from the Mellin transform of the density
$g_1(u_1)$ of $u_1$. Even though the density $g_1(u_1)$ is unknown,
its Mellin transform is available from (\ref{Eu1h}) for $h=s-1$. Hence from
the unique inverse Mellin transform
\begin{equation*}
g_1(u_1)=u_1^{-1}\frac{1}{2\pi i}\int_L[E(u_1^h)]u_1^{-h}{\rm d}h
\end{equation*}
where $i=\sqrt{-1}$, $L$ is a suitable contour and $E(u_1^h)$ is given in (\ref{Eu1h}). But the structure in
(\ref{Eu1h}) is that of a Mellin transform of a G-function of the type $G_{p,p}^{p,0}(\cdot)$. Hence
\begin{align}\label{gfunction}
g_1(u_1)&=u_1^{-1}\left[\prod_{j=1}^p\frac{\Gamma(\gamma+\frac{q}{2}+\frac{\eta}{1-\alpha}+\frac{p+1}{2}-\frac{j-1}{2}}{\Gamma(\gamma+\frac{q}{2}-\frac{j-1}{2})}\right]\notag\\
&\times G_{p,p}^{p,0}\left[\big|_{\gamma+\frac{q}{2}-\frac{j-1}{2}, j=1,...,p}^{\frac{\eta}{1-\alpha}+\frac{p+1}{2}, j=1,...,p}\right], 0<u_1<1
\end{align}

For $\alpha >1$,  from (\ref{c2}), $u_2=[a(\alpha-1)]^pv^2$ is a product
of $p$ statistically independently distributed real scalar type-2
beta random variables and hence the density of $u_2$ can be written
in terms of a G-function of the type $G_{p,p}^{p,p}(\cdot)$. \\

 Similarly
for $\alpha\rightarrow 1$, $u_3=(a\eta)^pv^2$ is structurally a
product of $p$ statistically independently distributed real gamma
random variables and the density of $u_3$ can be written in terms of
a G-function of the type
$G_{0,p}^{p,0}(\cdot)$.\\

A form such as the one in (\ref{u_1}) is connected to the $\lambda-$criterion in the likelihood ratio principle of testing statistical hypothesis on the parameters of one or more multivariate Gaussian populations. The pathway model for $\alpha<1$ can be structurally identified with a constant multiple of the one-to-one function of the $\lambda-$criterion in many situations in multivariate statistical analogues.
\vskip.3cm
\noindent
\emph{4.3 \hskip.3cm Connection to likelihood ratio criteria}
\vskip.3cm Consider the case $\alpha<1$. From (\ref{Eu1h}) and (\ref{u_1}) one can
see a structural representation in the form of a product of $p$
statistically independently distributed type-1 beta random
variables. Such a structure can also arise from a determinant of the
type
\begin{equation}\label{lamb}
\lambda=\frac{|G_1|}{|G_1+G_2|}
\end{equation}
where $G_1$ and $G_2$ are independently distributed real
matrix-variate gamma variables with the same scale parameter matrix.
Such matrix representations are recently examined in Mathai (2007).
A particular case of a real matrix-variate gamma matrix is a Wishart
matrix. When $G_1$ and $G_2$ are independently distributed Wishart
matrices $\lambda$ in (\ref{lamb}) corresponds to the likelihood ratio test
criterion or a one-to-one function of it in multivariate statistical
analysis. A large number of test criteria based on the principle of
maximum likelihood have the structure in (\ref{lamb}), with a representation
as in (\ref{u_1}) in the null case or when the statistical hypothesis is
true. Distribution of the test statistic, when the null hypothesis
is true, is known as the null distribution. Thus, a large number of
null distributions, associated with the tests of hypotheses on the
parameters of one or more multivariate normal populations, have the
structure in (\ref{lamb}) and thus a large number of cases are covered by
our discussion above. Hence a direct link can be established between
the volume of a random
$p$-parallelotope and the $\lambda$-criterion in (\ref{lamb}).\\
\vskip.2cm
Various types of generalizations of the Dirichlet distribution are recently studied by Jacob et al. (2004, 2005) and Kurian et al. (2004). In these papers there are several theorems
characterizing or uniquely determining these generalized Dirichlet models through a product of
independently distributed real scalar type-1 beta random variables. In all these cases one can
establish a direct link from (\ref{u_1}) to (\ref{lamb}) providing explicit representations of the densities
in terms of generalized Dirichlet integrals. As byproducts, these can also produce several results
connecting G-functions and Dirichlet integrals. One such characterization theorem is recently
explored by Thomas et al. (2008).
\vskip.3cm
\noindent
\emph{4.4 \hskip.3cm Quadratic forms}
\vskip.3cm
 For $p=1$
and $q>1$, (\ref{v2}) gives
\begin{equation}\label{quadratic}
Q=(X-M)B(X-M)'
\end{equation}
the quadratic form.\\

 Current theory of quadratic form in random
variables is based on the assumption of a multivariate Gaussian
population. But from (\ref{matrixpathway}) we have a generalized quadratic form in
$X$, where $X$ has the density in (\ref{matrixpathway}), and the density of $Q$ is
available from the step in (\ref{1}) to (\ref{gfunction}). The theory of quadratic
form in random variables can be extended to the samples from the density in (\ref{matrixpathway}),
rather than confining the study to Gaussian population. For the
study of quadratic and bilinear forms see Mathai and Provost (1992),
and Mathai, Provost and Hayakawa (1995). When $p=1$ the quadratic
form in (\ref{quadratic}) can be given many interpretations. Let $B=V^{-1}$ where
$V$ is the covariance matrix in $X$, that is, $V={\rm
cov}(X)=E[(X-\mu)'(X-\mu)]$ where $X-\mu$ is $1\times q, q>1$. Since
$Y=V^{-\frac{1}{2}}(X-\mu)'$ is the standardized $X$,
$YY'=y_1^2+...+y_q^2=(X-\mu)V^{-1}(X-\mu)'$ is the square of the
generalized distance between $X$ and $\mu$. Also
$(X-\mu)V^{-1}(X-\mu)'=c>0 $ is the ellipsoid of concentration in
$X$ or a scalar measure of the extent of dispersion in $X-\mu$. The
components in $X$ can be given physical interpretations. Consider a
growing and moving rain droplet in a tropical cloud. Then $x_1$
could be the surface area, $x_2$ could be the energy content, $x_3$
the velocity in a certain direction and so on. The components in
this case have joint variations. When $p=1$ one can derive the
pathway density in (\ref{matrixpathway}) through an optimization of a certain measure
of entropy also.
\vskip.3cm
\noindent
\emph{4.5 \hskip.3cm Pathway fractional integral}
\vskip.3cm
Recently, an extension of classical fractional integral operators of scalar functions of scalar variables
to the matrix-variate cases has been given by Mathai (2009). Real-valued scalar functions of
matrix argument, where the argument matrix is real and positive definite, are used in the extensions.
In this regard, a matrix-variate pathway fractional integral operator is introduced, see Seema Nair (2011)
which may be regarded as a generalization of matrix-variate Riemann-Liouville fractional integral
operator. Moreover, from this operator one can figure out all the matrix-variate fractional integrals and almost all the extended densities for the pathway parameter $\alpha<1$ and $\alpha \rightarrow 1$. Through this new fractional integral operator, one can go to matrix-variate gamma to matrix-variate Gaussian or normal density with appropriate parametric values. In the present paper we bring out the idea of matrix-variate pathway to the corresponding fractional integral transform. Consequently a scalar version of pathway fractional integral operator can also be deduced, which generalizes the classical Reimann-Liouville fractional integration
 operator. The pathway fractional integral operator has found applications in reaction-diffusion problems, non-extensive statistical mechanics, non-linear waves, fractional differential equations, non-stable neighborhoods of physical system etc.\\


The following definition and notation is given for the matrix-variate pathway fractional integral:
\begin{equation}\label{eqnp7}
(P_{O+}^{(\eta,\alpha)}f)(X)=|X|^{\eta-\frac{p+1}{2}}\int_T
|I-a(1-\alpha)X^{-\frac{1}{2}}T X^{-\frac{1}{2}}|^{\frac{\eta}{(1-\alpha)}-\frac{p+1}{2}}f(T) {\rm d}T,
\end{equation}
where $O$ stands for a null matrix, and $T=T'>0$ and $O$ are $p \times p$ matrices and $f(T)$ is a real-valued integrable scalar function of $T$ and ${\rm d}T$ is the wedge product of differentials. Also $a,\alpha$ scalars, $\eta \in C$ $a>0,~I-a(1-\alpha)X^{-\frac{1}{2}}T X^{-\frac{1}{2}}>0$ (positive definite) and $\alpha$ is the pathway parameter, $\alpha<1$. It is hoped that the matrix-variate extensions of the operator will enable researchers working in physical, chemical and engineering sciences to extend their theories to the
corresponding matrix-variate situations.\\

\noindent
 When $\alpha\rightarrow 1_{-}$,
$|I-a(1-\alpha)X^{-\frac{1}{2}}T X^{-\frac{1}{2}}|^{\frac{\eta}{(1-\alpha)}-\frac{P+1}{2}}\rightarrow {\rm
e}^{-a \eta ~\text{tr}[X^{-\frac{1}{2}}T X^{-\frac{1}{2}}]}$. This follows from the following facts: Since $X^{-\frac{1}{2}}T X^{-\frac{1}{2}}$ is real symmetric there exists an orthogonal matrix $Q$ such that $Q'X^{-\frac{1}{2}}T X^{-\frac{1}{2}}Q$ will be a diagonal matrix with eigenvalues being the diagonal elements. Then when the limit is applied, each factor goes to the exponent and sum up to equate to the trace. Thus the operator will become
\begin{eqnarray}\label{eqnp8}
(P_{O+}^{(\eta,1)}f)(X)&=&
|X|^{\eta-\frac{P+1}{2}} \int_{T=T'>0}{\rm
e}^{-\text{tr}[a \eta X^{-1}T]}~f(T)~{\rm d}T =|X|^{\eta-\frac{P+1}{2}}L_f(a \eta X^{-1}).\nonumber
\end{eqnarray}
When $\alpha=0,~a=1$ in (\ref{eqnp7}), the integral will become,
\begin{eqnarray}\label{eqnp9}
(P_{O+}^{(\eta,0)}f)(X)&=&|X|^{\eta-\frac{P+1}{2}}\int_{O<T<X}
|I-X^{-\frac{1}{2}}T X^{-\frac{1}{2}}|^{\eta-\frac{P+1}{2}}f(T){\rm d}T,~\Re(\eta)>\frac{p-1}{2}\nonumber\\
&=&\int_{O<T<X}
|X-T|^{\eta-\frac{P+1}{2}}f(T){\rm d}T=\Gamma_p(\eta)~{_OD_X^{-\eta}}f
\end{eqnarray}
where ${_OD_X^{-\eta}}$ is the matrix-variate extension of the standard left-sided Reimann-Liouville fractional
integral and is defined as
\begin{equation}\label{eqnp10}
{_OD_X^{-\eta}}f=\frac{1}{\Gamma_p(\eta)}\int_{O<T<X}
|X-T|^{\eta-\frac{P+1}{2}}f(T){\rm d}T,~~\Re(\eta)>\frac{p-1}{2}.
\end{equation}
When $p=1$ in (\ref{eqnp7}), one can obtain the scalar version of the pathway fractional integral operator, see Seema Nair (2009) and is defined by the following:\\

\noindent
Let $f(x)\in L(a,b),~ \eta \in C,~~\Re(\eta )>0,~a>0 $ and
$\alpha<1$, then
\begin{equation}\label{eqnp12}
(P_{0+}^{(\eta,\alpha)}f)(x)=x^{\eta-1}\int_0^{[\frac{x}{a(1-\alpha)}]}
[1-\frac{a(1-\alpha)t}{x}]^{\frac{\eta}{(1-\alpha)}-1}f(t) {\rm d}t,
\end{equation}
where $\alpha$ is the pathway parameter and $f(t)$ is an arbitrary
function. In Seema Nair (2009), it is shown that as $\alpha\rightarrow 1_{-}$, (\ref{eqnp12}) takes the form of Laplace transform of the arbitrary function $f(t)$. Again when $\alpha=0,~a=1$, (\ref{eqnp12}) reduces to the left-sided Reimann-Liouville fractional integral operator. By the way an idea of thicker or thinner tail model associated with Mittag-Leffler function is obtained. As a result, generalized gamma Mittag-Leffler density can be obtained as a limiting case of pathway operator. It is also shown that under the conditions $\alpha=0,~a=1$ and as $f(t)$ changes to $~_2F_1(\eta+\beta,~-\gamma;\eta;1-\frac{t}{x})f(t)$, (\ref{eqnp12}) yields the Saigo fractional integral operator. Thus we
can obtain all the generalizations, like Saigo (1978),
Erd\'{e}lyi-Kober [(1954), (1940)], etc., of left-sided fractional integrals by
suitable substitutions, so that we call it the pathway fractional operator, a path through $\alpha$,
 leading to the above known fractional operators.\\

The importance of the operator is that a connection is established to wide classes of statistical distributions, to several types of situations in physics, chemistry, to the input-output situations in social sciences, in reaction-diffusion problems etc. The pathway parameter $\alpha$ establishes a path of going from one family of distributions
to another family and to different classes of distributions. Thus, the pathway fractional operator, will enable us to derive a number of results covering wide range of distributions. The ``fractional integration'' nature
of the operator will then extend the corresponding results to wider ranges,
where when the pathway parameter $\alpha$ goes to 1 the corresponding results
on generalized gamma type functions are obtained.\\
\vskip.3cm
\noindent
{\bf{Acknowledgements}}
\vskip.2cm
\noindent Authors acknowledge gratefully the encouragement given by Professor A. M. Mathai, Department of Mathematics and Statistics, McGill University, Montreal,
Canada H3A 2K6.
\vskip.3cm
 \noindent
{\bf{References}}\vskip.3cm
{\scriptsize{
\noindent
Anderson, W. J., Haubold, H. J. $\&$ Mathai A. M. (1994). Astrophysical thermonuclear functions. {\em Astrophysics and Space $~~~~~$Science}, 214, 49-70.\\
\noindent
Beck, C. (2006). Stretched exponentials from superstatistics. {\em Physica
A}, {\bf 365}, 96-101.\\
\noindent
 Beck, C. $\&$ Cohen, E.G.D. (2003). Superstatistics. {\em Physica A},  322, 267-275.\\
\noindent Erd\'{e}lyi, A., Magnus, W., Oberhettinger F. and  Tricomi, F.G. (1953). {\em Higher Transcendental Functions},  Vol.1, McGraw-Hill, $~~~~~$New York.\\
\noindent
Haubold, H. J. $\&$ John, R. W. (1978). On the evaluation of an integral connected with the thermonuclear reaction rate in $~~~~~$the closed form. {\em Astronomische Nachrichten}, 299, 225-232.\\
\noindent
Haubold, H.J. \& Mathai, A.M. (1995). A heuristic remark on the periodic
variation in the number of solar neutrinos $~~~~~$detected on Earth. {\em Astrophysics
and Space Science}, 228, 113-134.\\
\noindent
Haubold, H.J. \& Mathai, A.M. (2000). The fractional kinetic equation and
thermonuclear functions. \emph{Astrophysics and $~~~~~$Space Science}, 273, 53-63.\\
\noindent
Haubold, H.J. \& Kumar, D. (2008). Extension of thermonuclear functions through
the pathway model including Maxwell-$~~~~~$Boltzmann and Tsallis distributions. {\em Astroparticle
Physics}, \textbf{29}, 70-76.\\
\noindent
Haubold, H.J., Kumar, D., Nair, S.S. \& Joseph, D.P. (2010). Special functions and
pathways for problems in astrophysics: $~~~~~$An essay in honor of A.M. Mathai. {\em Fractional
Calculus \& Applied Analysis}, \texttt{13}, 133-158.\\
\noindent
 Honerkamp, J. (1994). {\it Stochastic Dynamical Systems: Concepts, Numerical
Methods, Data Analysis.} VCH Publishers, $~~~~~$New York.\\
\noindent
Jacob, Joy; Jose, K.K. \& Mathai, A.M. (2004). Some properties of real
matrix-variate inverted generalized Dirichlet $~~~~~$integral. \emph{Journal of the Indian
Academy of Mathematics}, \textbf{26(1)}, 175-189.\\
\noindent
Jacob, Joy, Sebastian George \& Mathai, A.M. (2005). Some properties
of complex matrix-variate generalized Dirichlet $~~~~~$integrals.\emph{ Proceedings of the
Indian Academy of Sciences}, \textbf{15(3)}, 1-9.\\
\noindent
Joseph, D. P. (2009). Gamma distribution and extensions by using pathway
idea. {\it Statistical Papers} (accepted), DOI $~~~~~$10.1007/s00362-009-0231-y.\\
\noindent
Joseph, D. P. \& Haubold, H. J. (2009). Extended reaction rate
integral as solutions of some general differential equations. {\it
$~~~~~$Astrophysics and Space Science} (Proceedings), 41-52.\\
\noindent
Kilbas, A.A. \& Kumar, D. (2009). On generalized Kr¨atzel functions. \emph{Integral transforms and
Special functions}, \textbf{20(11)}, $~~~~~$836-845.\\
\noindent
Kilbas, A. A. \& Shlapakov, S. A. (1993). On Bessel-type integral
 transformation and its compositions with integral and $~~~~~$differential
 operators. \emph{Dokl. Akad. Nauk Belarusi}, \textbf {(37)4}, 10-14.\\
\noindent
Kilbas, A.A., Saxena, R.K. \& Trujillo, J.J. (2006). Kr\"{a}tzel
function as a function of hypergeometric type. \emph{Fractional $~~~~~$Calculus and Applied Analysis}, \textbf{(9)2},
109-130.\\
\noindent
Kober, H. (1940). On fractional integrals and derivatives. \emph{Quar. J. Math., Oxford Series}, II, 193-211.\\
\noindent
Kr\"{a}tzel, E. (1979). Integral transformations of Bessel type.
\emph{Generalized functions and Operational Calculus, Proc. Conf. $~~~~~$Varna,
Bulg. Acad. Sci. Sofia}, 148-465.\\
\noindent
Kumar, D. (2009). Type-2 P-transforms. \emph{Proceedings of AMADE-09}, Minsk, Belarus.\\
\noindent
Kurian, K.M., Kurian, Benny; \& Mathai, A.M. (2004). A matrix-variate
extension of inverted Dirichlet integral. \emph{Proceed-$~~~~~$ings of the National Academy
of Sciences (India)}, \textbf{74(A)II}, 1-10.\\
\noindent
 Mathai, A.M. (1993). {\it A Handbook of Generalized Special Functions for
Statistical and Physical Sciences}. Clarendon Press, $~~~~~$Oxford.\\
\noindent
Mathai, A.M. (1993a). On non-central generalized Laplacianness of
quadratic forms in normal variables. \emph{Journal of $~~~~~$Multivariate Analysis,} \textbf{45},
239-246.\\
\noindent
Mathai, A.M. (1993b). The residual effect of a growth-decay mechanism
and the distributions of covariance structures. \emph{The $~~~~~$Canadian Journal of
Statistics,} \textbf{21(3)}, 277-283.\\
\noindent
Mathai, A.M. (1997). \emph{Jacobians of Matrix Transformations and Functions
of Matrix Argument,} World Scientific Publishing, $~~~~~$New York.\\
\noindent
 Mathai, A.M. (2005). A Pathway to matrix-variate gamma and
 normal densities. {\em Linear Algebra and Its Applications}, {\bf{396}}, $~~~~~$317--328.\\
  \noindent
  Mathai, A.M. (2007). Random volumes under a general matrix-variate
model, Linear Algebra and Its Applications, \textbf{425}, $~~~~~$162-170.\\
  \noindent Mathai, A.M. (2009). Fractional integrals in the matrix-variate cases and connection to statistical distributions. \emph{Integral $~~~~~$Transforms and Special Functions}, {\bf 20(12)}, 871-882.\\
  \noindent
  Mathai, A. M. (2010). Some properties of Mittag- Leffler functions and matrix variate analogues: a statistical perspective. {\it{$~~~~~$Fractional Calculus \& Applied Analysis,}} \textbf{30(2)}, 113-132.\\
\noindent
 Mathai, A.M. $\&$ Haubold, H.J (1988). {\it Modern Problems in
Nuclear and Neutrino Astrophysics}, Akademic-Verlag Berlin.\\
\noindent
 Mathai, A.M. $\&$ Haubold, H.J (2002). Review of Mathematical techniques applicable in astrophysical reaction rate theory. {\em $~~~~~$Astrophysics and Space Science}, \textbf{282}, 265-280.\\
\noindent
 Mathai,  A.M. $\&$ Haubold,  H.J. (2007).  Pathway
model Pathway model, superstatistics Tsallis statistics and a
generalized $~~~~~$measure of entropy. \emph{Physica A},
\textbf{375}, 110-122.\\
\noindent
Mathai,  A.M. $\&$ Haubold,  H.J. (2007a). On generalized entropy measures and pathways. \emph{Physica A: Statistical Mechanics $~~~~~$and its Applications}.
{\bf 385}, 493-500.\\
\noindent
  Mathai,  A.M. $\&$ Haubold,  H.J. (2008). Pathway parameter and thermonuclear functions. \emph {Physica A: Statistical Mechanics $~~~~~$and its Applications} {\bf 387}, 2462-2470.\\
\noindent
 Mathai,  A.M. $\&$ Haubold,  H.J. (2010). A pathway from bayesian statistical analysis
to superstatistics. {\emph arXiv:1011.5658v1 $~~~~~$[cond-mat.stat-mech] 25 Nov 2010
A}.\\
\noindent
Mathai, A.M. \& Provost, S.B. (1992). \emph{Quadratic Forms in Random
Variables: Theory and Applications}, Marcel Dekker, $~~~~~$New York.\\
\noindent
Mathai, A.M. \& Provost, S.B. (2005). Some complex matrix-variate
statistical distributions on rectangular matrices. \emph{$~~~~~$Linear Algebra and Its Applications,}
\textbf{410}, 198-216.\\
\noindent
Mathai, A.M., Provost, S.B. \& Hayakawa, T. (1995). \emph{Bilinear Forms
and Zonal Polynomials, Lecture Notes}, Springer-$~~~~~$Verlag, New York.\\
  \noindent Mathai, A.M.,  Saxena, R.K. \&  Haubold, H. J. (2010). {\it The $H$-function Theory and Applications},
  Springer, New York, $~~~~~$London
 and Sidney.\\
\noindent
  Mathai, A.M.  $\&$  Rathie, P.N.  (1975). \emph{Basic Concepts in Information
 Theory and Statistics: Axiomatic Foundations and $~~~~~$Applications}, Wiley Halsted,
 New York and Wiley Eastern, New Delhi.\\
\noindent
 Pillai, R.N. (1990). On Mittag-Leffler functions and related distributions. {\em Ann. Inst. Statist. Math.}, {\bf 42}, 157-161.\\
\noindent
Saigo, M. (1978). A remark on integral operators involving the Gauss hypergeometric functions. \emph{Math. Rep. Kyushu Univ.}, $~~~~~$\textbf{11}, 135-143.\\
\noindent
Sebastian, N. (2009). A generalized gamma model associated with
Bessel function and its applications in statistical $~~~~~$mechanics.
\emph{Proceedings of AMADE-09}, Proceedings of Institute of
Mathematics, Belarus.\\
\noindent
Sebastian, N. (2011). A generalized multivariate gamma model
associated with Bessel function. \emph{Integral transforms and
$~~~~~$Special functions}, \textbf{22(9)}, 631--645.\\
\noindent
 Seema Nair, S., (2009). Pathway fractional integration operator. \emph{Fractional Calculus and Applied Analysis}, {\bf{12(3)}}, 237--252.\\
\noindent
 Seema Nair, S., (2011). Pathway fractional integral operator and matrix-variate
functions. {\em Integral Transforms and Special $~~~~~$Functions}, {\bf 22(3)}, 233-244.\\
\noindent
 Seema Nair, S. and Kattuveettil, A. (2010). Some remarks on the paper "On the $q$-type distributions" . {\it
Astrophysics and $~~~~~$Space Science} (Proceedings), 11-15.\\
\noindent
Shanoja Naik, R. (2008). \emph{Pathway Distributions, Autoregressive Processes and their Applications}, Thesis.\\
\noindent
Thomas, S., Thannippara, A. and Mathai, A.M. (2008). On a matrix-variate generalized type-2 Dirichlet model. \emph{Advances $~~~~~$and Applications in Statistics}, \textbf{8(1)}, 37-56.\\
\noindent
 Tsallis, C. (1988). Possible generalizations
of Boltzmann-Gibbs statistics. {\it{Journal of Statistical
Physics}}, {\bf{52}}, 479-487.\\
\noindent
Tsallis, C. (2004). What should a statistical mechanics satisfy to reflect
nature? {\it {Physica D}}, {\bf 193}, 3-34.

\end{document}